\documentclass[fleqn,12pt,twoside]{article}
\usepackage{espcrc1,epsfig}

\def\Journal#1#2#3#4{{#1} {#2} (#4) #3}

\newcommand\RMP{Rev. Mod. Phys.}

\newcommand\PR{Phys. Rev.}

\newcommand\EPC{{Eur. Phys. J.} C}
\newcommand\NIMA{Nucl. Inst. and Meth. A}
\newcommand\NPS{Nucl. Phys. B (Proc. Suppl.)}
\newcommand\NPB{ Nucl. Phys. B}
\newcommand\PLB{Phys. Lett.  B}
\newcommand\PRL{Phys. Rev. Lett.}
\newcommand\PRD{Phys. Rev. D}

\newcommand\ea{{\em et al.}}

\newcommand{\amu}[1][]{\ensuremath{a_{\mu^{#1}}}}
\newcommand{\gm}{\ensuremath{(g-2)}}
\newcommand{\wa}{\mbox{\ensuremath{\omega_a}}}
\renewcommand{\wp}{\mbox{\ensuremath{\omega_p}}}

\title{The Muon Anomalous Magnetic Moment and the Standard
Model}

\author{David W. Hertzog
\address{Department of Physics, University of Illinois
at Urbana-Champaign \\ Urbana, IL 61801, USA} \\ {\em on behalf of
the Muon \gm\ Collaboration}\thanks{E821 Collaboration: Boston:
R.M.~Carey, E.~Efstathiadis, M.F.~Hare, X.~Huang, F.~Krienen,
A.~Lam, I.~Logashenko, J.P.~Miller, J.~Paley, Q.~Peng, O.~Rind,
B.L.~Roberts, L.R.~Sulak, A.~Trofimov; BNL: G.W.~Bennett,
H.N.~Brown, G.~Bunce, G.T.~Danby, R.~Larsen, Y.Y.~Lee, W.~Meng,
J.~Mi, W.M.~Morse, D.~Nikas, C.~\"Ozben, R.~Prigl,
Y.K.~Semertzidis, D.~Warburton; Cornell U.: Y.~Orlov; Heidelberg:
A.~Grossmann, G.~zu~Putlitz, P.~von~Walter; Illinois:
P.T.~Debevec, W.~Deninger, F.E.~Gray, D.W.~Hertzog,
C.J.G.~Onderwater, C.~Polly, M.~Sossong, D.~Urner; KEK:
A.~Yamamoto; KVI: K.~Jungmann; Minnesota:  P.~Cushman, L.~Duong,
S.~Giron, J.~Kindem, I.~Kronkvist, R.~McNabb, T.~Qian, P.~Shagin;
Novosibirsk:  V.P.~Druzhinin, G.V.~Fedotovich, D.~Grigoriev,
B.I.~Khazin, N.M.~Ryskulov, Yu.M.~Shatunov, E.~Solodov; Tokyo:
M.~Iwasaki; Yale: H.~Deng, S.K.~Dhawan, F.J.M.~Farley,
V.W.~Hughes, D.~Kawall, M.~Grosse~Perdekamp, J.~Pretz, S.I.~Redin,
E.~Sichtermann, A.~Steinmetz.} }

\begin{document}

\maketitle

\begin{abstract}
The muon anomalous magnetic moment measurement, when compared with
theory, can be used to test many extensions to the standard model.
The most recent measurement made by the Brookhaven E821
Collaboration reduces the uncertainty on the world average of
$a_{\mu}$ to $0.7$~ppm, comparable in precision to theory. This
paper describes the experiment and the current theoretical efforts
to establish a correct standard model reference value for the muon
anomaly.
\end{abstract}

\section{INTRODUCTION}

\begin{figure}
\begin{center}
\psfig{figure=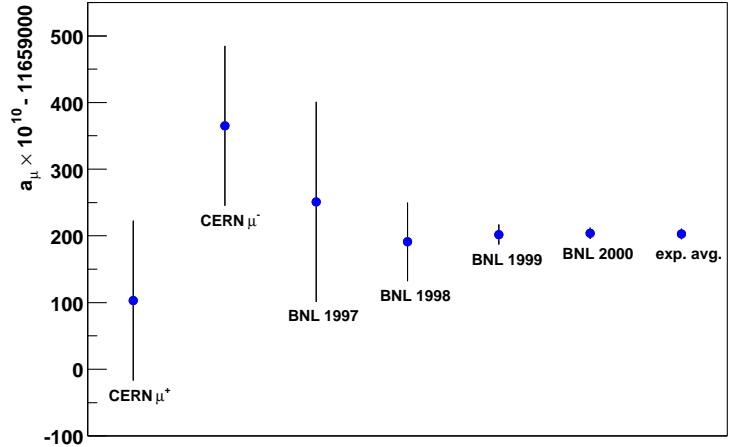,width=0.6\textwidth}
\caption{Progression of precision in the measurement of the muon
anomalous magnetic moment.  The current world average has an
uncertainty of 0.7~ppm.
Refs.~\cite{cernIII,carey99,brown00,brown01,bennett02}
\label{fg:exp-results}}
\end{center}
\end{figure}

A precision measurement of the muon anomalous magnetic moment,
$a_{\mu}~= \gm/2$, tests the standard model to the extent that
both theory and experiment are well understood.  The BNL E821
experiment has steadily increased the precision with which $\amu$
is known~\cite{carey99,brown00,brown01,bennett02}. The result,
published~\cite{bennett02} in August 2002 dominates the new world
average of $\amu(exp)~=~11~659~203(8)\times 10^{-10}$ (0.7~ppm).
This includes the complete analysis of all of the positive muon
data obtained to date by the E821 group. Another (somewhat
smaller) sample of negative muon data is presently being analyzed,
and a final run to complete the experiment is planned.
Figure~\ref{fg:exp-results} illustrates the progression of
precision in the measurements of $\amu$ including the
CERN-III~\cite{cernIII} and present BNL results.

The standard model (SM) expectation for \amu~ is based on QED, the
weak interaction, and hadronic loops.
All must be computed at the sub-ppm level in order to be compared
meaningfully with experiment. Of these contributions, both the QED
and weak terms are believed to be known to sufficient
precision~(see Czarnecki and Marciano~\cite{marciano99}), even
withstanding a recent examination of eighth-order QED terms, which
uncovered a negligible mistake~\cite{kin02}, and considerations of
hadronic contributions to higher-order weak terms~\cite{knect02}.
The need to consider such subtle contributions has reached
increased importance because the BNL experimental precision is
better than 1~ppm, additional data is being analyzed, and
Collaboration discussions have begun to consider an even more
precise experiment, either at BNL or the JHF. Many standard model
extensions manifest themselves in additional contributions to
$\amu$ at the ppm level, so the relevance of the current studies,
in advance of the LHC direct-discovery potential, is quite
important.

This paper briefly summarizes the experiment, the analysis of the
2000 data, and the up-to-date theory ({\em i.e.}, through
Nov.~2002). The reader is strongly encouraged to consult the
original published papers where appropriate.

\subsection{Principle of the Experiment}

The experiment measures directly \gm~ and not $g$, where $\gm/2$
is approximately $\alpha/2\pi \approx 1/860$. The muon anomaly is
determined from the difference between the cyclotron and spin
precession frequencies for muons contained in a magnetic storage
ring, namely
\begin{equation}
   \vec \omega_a = - {e \over m }\left[ a_{\mu} \vec B -
   \left( a_{\mu}- {1 \over \gamma^2 - 1}\right)
   \vec \beta  \times \vec E \right].
   \label{eq:omega}
\end{equation}

Because electric quadrupoles are used to provide vertical
focussing in the storage ring, the $\vec{\beta}  \times \vec{E}$
 term is necessary and illustrates the sensitivity of the spin motion to
a radial electric field. This term conveniently vanishes for the
``magic" momentum of 3.094~GeV/$c$ when $\gamma = 29.3$. The
experiment is therefore built around the principle of production
and storage of muons centered at this momentum in order to
minimize the electric field effect.  Because of the finite
momentum spread of the stored muons, a 0.3~ppm correction to the
observed precession frequency is made to account for the muons
above or below the magic momentum. Vertical betatron oscillations
induced by the focussing electric field imply that the plane of
the muon precession has a time-dependent pitch. Accounting for
both of these electric-field related effects introduces a
$+0.76(3)$~ppm correction  to the precession frequency in the 2000
data.

Muons exhibit circular motion in the storage ring and their spins
precess until the time of decay ($\gamma\tau_{\mu}\approx 64.4~\mu
s$). The net spin precession depends on the integrated magnetic
field experienced along the path of the muon. Parity violation
leads to a preference for the highest-energy decay electrons (in
the muon rest frame) to be emitted in the direction of the muon
spin. Therefore, a snapshot of the muon spin direction at time $t$
after injection into the storage ring is obtained, again on
average, by the selection of decay electrons in the upper part of
the Lorentz-boosted Michel spectrum ($E_{max}\approx 3.1$~GeV).
The number of electrons above a selected energy threshold is
modulated at frequency $\wa$ with a threshold-dependent asymmetry
$A=A(E_{th})$. The decay electron distribution is described by
\begin{equation}
N(t) = N_{0}\exp(-t/\gamma\tau_{\mu})\left[1 + A\cos(\wa t +
\phi)\right], \label{eq:fivepar}
\end{equation}
where $N_{0}$, the normalization, $A$ and $\phi$ are all dependent
on the energy threshold $E_{th}$. For $E_{th} = 2.0$~GeV, $A
\approx 0.4$.

In summary, the experiment involves the measurement of the
time-averaged quantities $\wa$ and $B$ in Eq.~\ref{eq:omega}. Here
$B$ implies the azimuthially-averaged magnetic field, folded with
the muon distribution, the latter being determined from the
debunching rate of the initial beam burst and from a tracking
simulation. The field strength is measured using nuclear magnetic
resonance (NMR) in units of the free proton precession frequency,
$\omega_{p}$. The muon anomaly is computed from
\begin{equation}
a_{\mu} = \frac{\wa/\wp}{\lambda -\wa/\wp}\label{eq:actual},
\end{equation}
where $\lambda$ is the measured~\cite{pdg} magnetic moment ratio
$\mu_{\mu}/\mu_{p}=3.183~345~39(10)$.

Just like in the analysis of our 1999 data, four independent teams
evaluated $\wa$ and two studied $\wp$. During the analysis period,
the $\wa$ and $\wp$ teams maintained separate, secret offsets to
their measured frequencies. The offsets were removed and $a_{\mu}$
was computed only after all analyses were complete.

\section{EXPERIMENT}
\subsection{Storage Ring and Magnetic Field Measurement}
The Brookhaven muon storage ring~\cite{danby} is a superferric
``C"-shaped magnet, 7.11~m in radius, and open on the inside.
Three superconducting coils provide a central field of
approximately 1.45~T. Passive and active shimming elements have
been improved and optimized since the inception of the experiment.

The magnetic field is measured using pulsed NMR on protons in
water- or Vaseline-filled probes~\cite{prigl}. The proton
precession frequency is proportional to the local field strength
and is measured with respect to the same clock system employed in
the determination of \wa.  The absolute field is, in turn,
determined by comparison with a precision measurement of
$\omega_p$ in a spherical water sample~\cite{NMRabsolute} to a
relative precision of better than $10^{-7}$. A subset of the 360
``fixed" probes is used to continuously monitor the field during
data taking. These probes are embedded in machined grooves in the
outer upper and lower plates of the aluminum vacuum chamber and
consequently measure the field just outside of the actual storage
volume. Constant field strength is maintained using 36 of the
probes in a feedback loop with the main magnet power supply.

\begin{figure}
\begin{center}
\psfig{figure=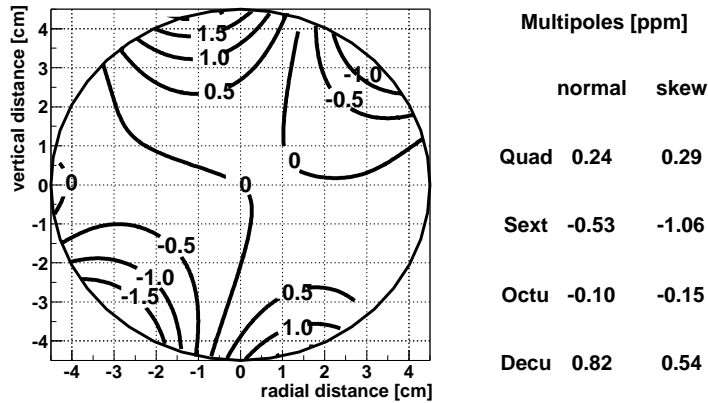,width=0.6\textwidth}
\caption{Two-dimensional multipole expansion, averaged over
azimuth for the 2000 data-taking period.  Half-ppm contours are
shown with respect to a central value of 1.451274~T. Multipole
amplitudes, relative to this field, are given at the beam
aperture. \label{fg:field2000}}
\end{center}
\end{figure}

The determination of the field inside the storage volume is made
by use of a unique non-magnetic trolley which can travel in vacuum
through the muon storage volume. The trolley carries 17 NMR probes
on a grid appropriate to determine the local multipolarity of the
field versus azimuth. Trolley field maps are made every few days
and take several hours to complete. Figure~\ref{fg:field2000}
shows the contours superimposed on a 9~cm diameter circle, which
represents the storage aperature.

\subsection{Creating and Storing Muons}
A highly polarized ($\approx 95\%$) muon sample is created from
the decay of 3.15~GeV/$c$ pions in a 72~m long decay channel. The
last dipole in this channel is tuned to a momentum approximately
1.7~percent lower than the central pion momentum in order to
enhance the muon fraction in the beam, compared to pions, at the
entrance to the storage ring. The approximately 1:1 $\pi$ to $\mu$
mix passes through a superconducting inflector
magnet~\cite{krienen}, which provides a field-free channel in the
back of the ring's iron yoke. A set of three pulsed ``kicker"
magnets~\cite{kickerNIM} provides a 10~mrad transverse deflection
to the muon bunch during the first two turns in the ring and thus
places the muons on a central orbit at the magic momentum.

Electric quadrupoles~\cite{quadsNIM} are initially charged
asymmetrically to scrape the beam against internal collimaters.
After approximately 20~$\mu$s, the voltages are symmetrized at
$\pm$24~kV corresponding to a weak-focussing field index
$n=0.137$. Coherent horizontal and vertical betatron oscillations
(CBO, i.e., oscillations of the beam as a whole), manifest
themselves as additional frequencies in the data that must be
accommodated (see below). The most important of these is the
beating between the cyclotron and the horizontal betatron
oscillations, determined as $f_{CBO} = (1-\sqrt{1-n})f_c =
466$~kHz, where $f_c$ is the cyclotron frequency.  Note that
$f_{CBO}$ is close to twice the precession frequency, $f_{g-2}
\approx 229$~kHz, and thus, if ignored, will affect the extraction
of $\omega_a$. This issue was handled in the 1999 data analysis by
accounting for the modulation of the overall rate of events at the
horizontal CBO frequency. However, for 2000, $n$ was adjusted
slightly, placing $f_{CBO}$ closer to $2f_{g-2}$ compared to 1999.
This effect, coupled with significantly higher statistics in 2000,
necessitated additional considerations. For example, not only was
the overall event rate seen to be modulated, but also the
asymmetry and even the phase. The four analyses approached this
central problem differently.

\subsection{Electron Detection and Pulse Fitting}
\begin{figure}
\begin{center}
{\psfig{figure=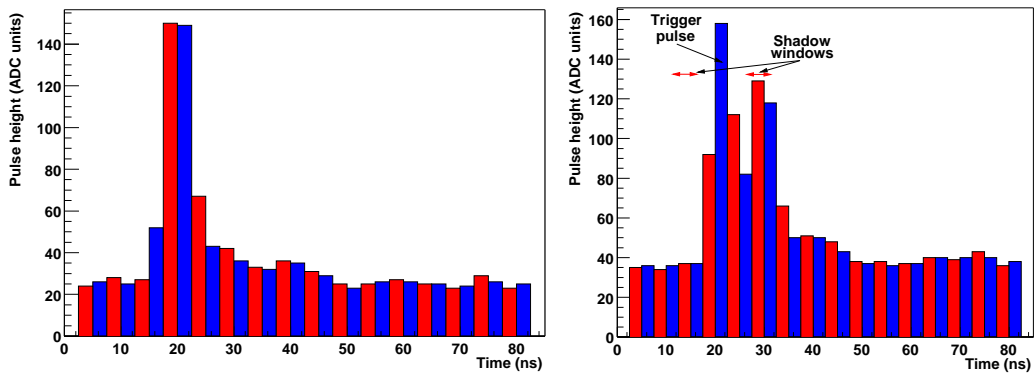,width=.9\textwidth}}
\caption{The left panel shows a single pulse on top of a pedestal.
The alternating shades illustrate the two independent samples of
the WFD. The right panel shows a more complicated pattern, in
which two pulses appear close together in time. \label{fg:pulses}}
\end{center}
\end{figure}
The electron detection system consists of 24 lead-scintillating
fiber electromagnetic calorimeters~\cite{sedykh} located
symmetrically around the inside of the storage ring and placed
immediately adjacent to the vacuum chamber.  The 23~cm long,
radially-oriented fiber grid terminates on four lightguides which
pipe the light to independent Hamamatsu R1828 2-inch PMTs.  The
PMT gains are carefully balanced because the four analog signals
are added prior to sampling by a dual-phase 400~MHz waveform
digitizer. At least 16 digitized samples (usually 24 or more),
making ``islands,'' are recorded for each decay electron event
exceeding a hardware threshold of approximately 1~GeV. Two
examples of such samples are shown in Fig.~\ref{fg:pulses}. The
left panel illustrates a simple, single pulse. Offline
pulse-finding and fitting techniques are used to establish the
electron energy and muon time of decay. Two quasi-independent
implementations of an algorithm to extract these quantities formed
the basis of two raw-data processing efforts. The methods differ
in particular for events which feature multi-pulse pileup, as
shown on the right panel in the figure. Here, two pulses are
actually sufficiently well-separated that both pulse-finding
algorithms can efficiently and accurately determine the correct
energies and times (verified by Monte Carlo simulations).

The pulse-finding algorithms also identify the extra events on the
tails of the recorded island of samples, which are then used to
estimate the time-dependent pileup fraction. These ``shadow
pulses'' are used to construct pileup-only event spectra, which
can be subtracted (on average) from the data.  This forms a
``pileup-free'' decay spectrum. Alternatively, the pileup-only
spectrum can be fit to determine the parameters necessary to
describe the distortion to the uncorrected decay spectrum.

\section{ANALYSIS OF $\omega_a$}
The data set for 2000 consisted of approximately four billion
electrons with an energy above 2.0~GeV, recorded $50~\mu$s or more
after muon injection.
These data came from a
common set of good runs that avoided known hardware failures,
glitches or calibration periods. Data obtained during systematic
studies were also removed. The four independent analysis methods
then treated the data quite differently after that.
\begin{figure}
\begin{center}
\psfig{figure=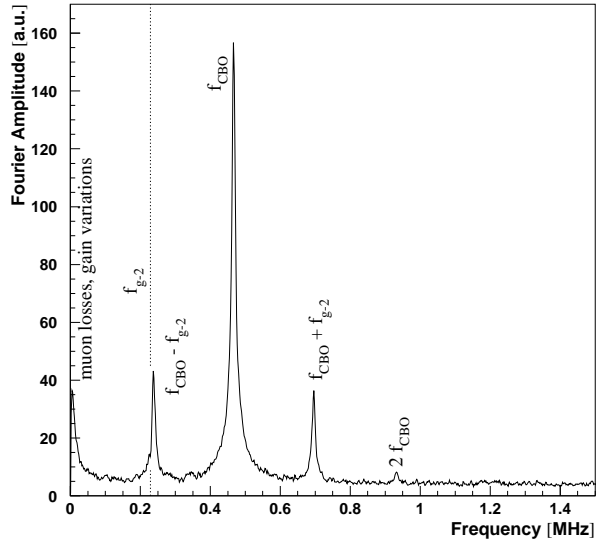,width=.5\textwidth} \caption{Fourier
transform of the residuals from the fit to the data using the
simple five-parameter function in Eq.~\ref{eq:fivepar}.
\label{fg:residuals}}
\end{center}
\end{figure}

A fit to the data using the five-parameter function given in
Eq.~\ref{eq:fivepar} is poor owing to the CBO modulation mentioned
above, the deviation from the pure exponential decay due to
time-dependent muon losses, and slowly-varying effects such as
detector gain or incomplete treatment of pileup.
Figure~\ref{fg:residuals} is a Fourier transform of the residuals
from such a simple fit. It clearly illustrates the dominant
horizontal CBO frequency and its sidebands at $f_{CBO}\pm
f_{g-2}$. Note that $f_{g-2}$ has been removed in this
representation because of the fit (the frequency is indicated by
the dashed line); the small peak at $f_{CBO}-f_{g-2}$ is close to
this dashed line and thus potentially interferes with the proper
extraction of $f_{g-2}$ or, equivalently, $\omega_a$.

Two analysis methods used independent implementations of an
expanded conventional functional form applied to a pileup-free
electron decay spectrum (pileup events were corrected for as
described above). The data were summed for all detectors, which
has the effect of reducing the CBO-related amplitudes
significantly compared to fits to individual detector spectra. To
account for the modulation of the normalization {\em and} the
asymmetry, $N_0\rightarrow N_{0}(t)$ and $A \rightarrow A(t)$ in
Eq.~\ref{eq:fivepar}.  The phase was left constant, relying
instead on the cancellation of its effect from summing all
detectors. The residual effect was included in the systematic
uncertainty associated with the overall treatment of CBO.

\begin{figure}
\begin{center}
\psfig{figure=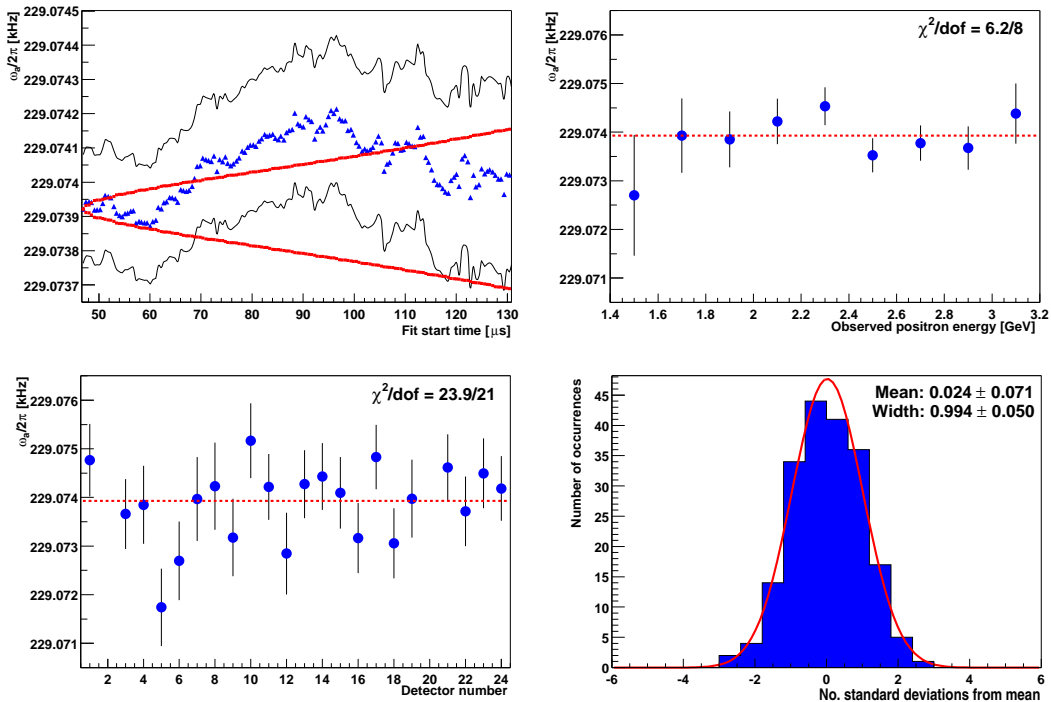,width=.9\textwidth}
\caption{Fit results from the energy-binned analysis versus start
time of the fit (upper left), energy bin (upper right), detector
station (lower left).  The distribution of the 198 individual
results with respect to the average is shown in the lower right
panel. \label{fg:finalresults}}
\end{center}
\end{figure}
A third analysis method started from a spectrum of the data where
the exponential decay and other slowly-varying effects are absent,
thus isolating the $(g-2)$ oscillation as only identifiable
feature.  Events are sorted into four sets ($A, B, C,$ and $D$),
with sets $A$ and $B$ time shifted by plus or minus one half of a
$g-2$ period, respectively, and sets $C$ and $D$ left unaffected.
Recombined in the ratio $(A-B)/(C+D)$, this spectrum exhibits an
oscillation about zero at frequency $\omega_a$. It was then fit
using a three-parameter function of the form $r(t) \approx
A(E)\sin(\omega_{a}t+\phi_{a}(E))$.

The fourth method divided the data into nine 0.2~GeV wide energy
bins for each of the 22 included detectors, resulting in 198
independent data sets.  The lowest-energy bin was centered at
1.5~GeV, which meant a significant amount of additional data was
considered compared to those analyses described above. The
lower-energy threshold also necessitated an extended pileup
identification procedure to account for events below twice the
hardware threshold. A pileup spectrum was built for each detector
and energy bin and fit to extract the amplitude and phase. These
parameters were then used in the fit function applied to the data,
which was not corrected for pileup, contrary to the other methods.
As in the first two methods, a slowly-varying muon loss function
was incorporated into the fit. Finally the terms $N_{0}, A,$ and
$\phi$ in Eq.~\ref{eq:fivepar} are permitted to oscillate with the
CBO frequency as $1 + A(t)_{i}\sin(\omega_{CBO}t+\phi_i)$, where
$A_i$ and $\phi_i$ are the relative amplitude and phase of the
modulation for $N_{0}, A,$ and $\phi$.  The modulation of these
parameters has a time-dependent envelope, which is determined both
from independent studies and from the fit itself.  Each detector
and energy bin was separately fit, with a start time determined
from a $\chi^{2}$ stability test.

Figure~\ref{fg:finalresults} illustrates the results on $\omega_a$
for the fourth method.  It shows the fit results versus start time
of the fit (upper left), energy bin (upper right), and detector
station (lower left).  The distribution of the 198 individual
results with respect to the average is shown in the bottom right
panel. All plots show the expected statistical fluctuations.

\subsection{Experimental Results}
Summarizing the four methods above involved accounting precisely
for data overlap, then combining to form a proper average.
Systematics were studied separately for each method.  The final
result for the 2000 precession frequency is $\omega_{a}/(2\pi) =
229~074.11(14)(7)$~Hz (0.7 ~ppm) where the first uncertainty is
statistical and the second systematic.  Combined with the field
value, and weighing with earlier experimental results, the new
result for the muon anomaly is $a_{\mu}(exp) = 11~659~203(8)
\times 10^{-10}$.

\subsection{Update on Theory}
The SM prediction for \amu~ has gone through several significant
changes during the last year; at present there is no single
quotable number. The QED~\cite{QED} and weak
contributions~\cite{marciano99,knect02} are not controversial at
the level of relevance required here giving, respectively,
$\amu$(QED)$~=~11~658~470.57(0.29)\times 10^{-10}$ (0.025 ppm) and
$\amu$(weak)$~=~15.1(0.4)\times 10^{-10}$ (0.03 ppm).

The first-order hadronic vacuum polarization (HVP) contribution
carries the largest uncertainty in the SM value.  It has been
updated in 2002 using new $e^{+}e^{-}$ data from Novosibirsk and
Beijing and additional hadronic tau decay data from LEP and CLEO.
These data are used as input into the dispersion relation to
compute $\amu$(HVP); see Ref.~\cite{DH98}. Davier
\ea~\cite{DEHZ02} (DEHZ) recently released an updated evaluation,
thus superseding their previous work and the compilations of
others (which were sometimes based on preliminary data). The most
important finding in DEHZ is that the $e^{+}e^{-}$ and tau-based
analyses do not agree with one another. The $e^{+}e^{-}$
evaluation was performed independently~\cite{hagiwara02}
confirming the result of DEHZ, but the difference with the
tau-based analysis appears to be more fundamental. For example,
the $\pi\pi$ spectral functions are not consistent and the
difference is shown to be energy dependent. Because the tau input
requires invoking CVC and isospin corrections, suspicion first
fell to it. These corrections are relatively
small~\cite{ciringliano02}, and even expanding the uncertainty
considerably would not put these data into agreement with the
$e^{+}e^{-}$-based result.

Meanwhile, the first reports on a third method to determine HVP
using the ``radiative return'' method have been
presented~\cite{kloe02}. Fixed center-of-mass energy $e^{+}e^{-}$
collisions are scanned for events having an initial-state radiated
photon, thus lowering the effective collision energy.  At one
accelerator setting, the entire spectral function can be obtained.
The reports are preliminary and use only a fraction of the data.
However, the procedure looks promising and should be able to
confirm or refute the $e^{+}e^{-}$ Novosibirsk result. For now,
DEHZ quote $\amu$(HVP,$e^{+}e^{-}$)$ = 684.7(7.0)$ or
$\amu$(HVP,tau)$ = 701.9(6.2)$, both $\times 10^{-10}$ with
comparable relative precisions of $\approx 0.6$ ppm. Because the
dispersion relation is weighed heavily toward low energies, the
true underlying difference in the input data exceeds the relative
difference implied to \amu.  One should not be tempted to take an
average or to ascribe the problem to statistical fluctuations.

Higher-order hadronic contributions~\cite{krause97} give
$\amu$(H-h.o.)$~=~-10.0(0.6)\times 10^{-10}$. The hadronic
light-by-light contribution changed by $200\%$ following Knecht
and Nyffeler's demonstration~\cite{knecht} that this contribution
must be positive, contrary to the existing
literature~\cite{hayakawa98,bijnens95} by others. Eventually
mistakes in earlier work were found~\cite{hayakawa01,bijnens01}
and agreement exists on the sign (with confidence), the magnitude
(essentially), but not necessarily the uncertainty (see, for
example, Ref.~\cite{musolf02}).  A middle-ground value with a
conservative uncertainty is $\amu$(H-LbL)~$=~8.6(3.2)\times
10^{-10}$.

Summarizing the above, we obtain the standard model theory to
date: $$ ~~~\amu(\rm{SM},e^{+}e^{-}{\rm
-based})~=~11~659~169.0(7.7)~(0.66~{\rm ppm)~or}$$ $$\amu({\rm SM,
tau-based})~=~11~659~186.2(7.0)~(0.60~{\rm ppm}).$$

\begin{figure}
\begin{center}
\psfig{figure=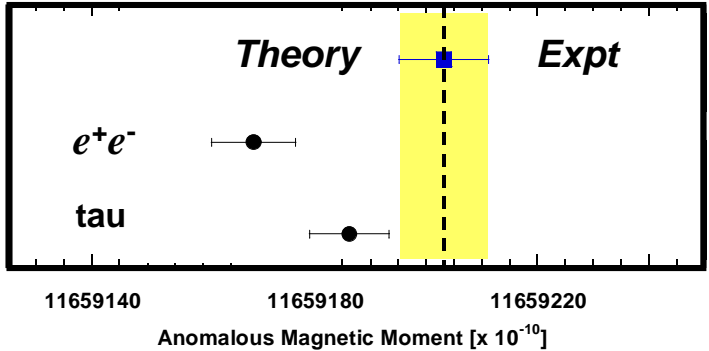,width=.70\textwidth}
\caption{Comparison of experiment with the standard model using
either $e^{+}e^{-}$-based or tau-based input for the HVP.
\label{fg:finalcomparison}}
\end{center}
\end{figure}

Compared to the experiment there is either a 3.0 ($e^{+}e^{-}$) or
a 1.6 (tau) standard deviation difference, respectively, leading
one to begin to draw very different conclusions (see
Fig.~\ref{fg:finalcomparison}).

The next year for $(g-2)$ might turn out to be as interesting as
the past year: a final data set will be analyzed from negative
muons; new input from KLOE and possibly BaBar should resolve the
HVP problem; and, finally, we hope to obtain the funds to complete
the experiment at BNL and reach the systematic limit of the
experiment.  Perhaps at that time, we can more meaningfully
explore the implications for physics beyond the standard model. At
present, the story is clearly still unfolding.

\section{Acknowledgments}
The \gm~ experiment is supported in part by the U.S. Department of
Energy, the U.S. National Science Foundation, the German
Bundesminister f\"{u}r Bildung und Forschung, the Russian Minsitry
of Science, and the US-Japan Agreement in High Energy Physics. The
author thanks his collaborators, and the organizers of PANIC'02
for an excellent meeting.


\begin{thebibliography}{9}
  \bibitem{carey99} R.M. Carey \ea, \Journal{\PRL}{82}{1632}{1999}.
  \bibitem{brown00} H.N. Brown \ea, Muon $\gm$ Collaboration,
                \Journal{\PRD}{62}{091101}{2000}.
  \bibitem{brown01} H.N. Brown \ea, \Journal{\PRL}{86}{2227}{2001}.
  \bibitem{bennett02} G.W. Bennett, \ea, Muon $\gm$ Collaboration,
                    \Journal{\PRL}{89}{101804}{2002}.
  \bibitem{cernIII}J. Bailey, \ea,  \Journal{\NPB}{150}{1}{1979}.
%
  \bibitem{marciano99}A.  Czarnecki and W.J.  Marciano, \Journal{\NPS}
                      {76}{245}{1999}.
  \bibitem{kin02}T. Kinoshita and M. Nio, hep-ph/0210322.
  \bibitem{knect02}M. Knecht, S. Peris, M. Perrottet and E. de
    Rafael, hep-ph/0205102

  \bibitem{pdg} D.E. Groom \ea, Review of Particle Physics,
  \Journal{\EPC}{15}{1}{2000}.
  \bibitem{danby} G.T. Danby \ea, \Journal{\NIMA}{457}{151}{2001}.
  \bibitem{prigl} R.  Prigl, U.  Haeberlen, K.  Jungmann, G.  zu
    Putlitz and P.  von Walter, \Journal{\NIMA}{374}{118}{1996}.
  \bibitem{NMRabsolute} W.D. Phillips \ea, Metrologia {\bf 13}, 81
    (1977); X.  Fei, V.W.  Hughes and R.  Prigl,
    \Journal{\NIMA}{394}{349}{1997}.
  \bibitem{krienen} F.  Krienen, D.  Loomba, and W.
                    Meng, \Journal{\NIMA}{283}{5}{1989}.
  \bibitem{kickerNIM}  E Efstathiadis \ea,
  \Journal{\NIMA}{496}{8}{2003}.

  \bibitem{quadsNIM}Y.K. Semertzidis \ea, Nucl. Instr. Methods A., in press.


  \bibitem{sedykh} S. Sedykh \ea, \Journal{\NIMA}{455}{346}{2000}.
 \bibitem{QED}P. Mohr and B. Taylor, \Journal{\RMP}{72}{351}{2000}.
 \bibitem{DH98}M. Davier and A. H\"ocker, \Journal{\PLB}{435}{427}{1998}.
 \bibitem{DEHZ02}M. Davier, S. Eidelman, A. H\"ocker and Z. Zhang, hep-ph/0208177.
  \bibitem{hagiwara02} K. Hagiwara, A.D. Martin, D. Nomura and T.
  Teubner, hep-ph/0209187.
  \bibitem{ciringliano02}V. Ciringliano, G. Ecker and H. Neufeld,
  hep-ph/0207310.

  \bibitem{kloe02}G. Vananzoni for the KLOE Collaboration, hep-ex/0210013.
  \bibitem{krause97}B. Krause, \Journal{\PR}{B390}{392}{1997};
    R.~Alemany, M.~Davier, A.~H\"ocker, \Journal{EPJ}{C2}{123}{1998}.
  \bibitem{knecht} M. Knecht and A. Nyffeler, \Journal{\PRD}{65}{073034}{2001},
    and M.~Knecht, A.~Nyffeler, M.~Perrottet, and E.~de~Rafael,
    \Journal{\PRL}{88}{071802}{2002}.
  \bibitem{hayakawa98} M. Hayakawa and T. Kinoshita,
    \Journal{\PRD}{57}{465}{1998}.
  \bibitem{bijnens95} J. Bijnens, E. Pallante, and J. Prades,
    \Journal{\PRL}{75}{1447}{1995}.
  \bibitem{hayakawa01}M. Hayakawa and T. Kinoshita, hep-ph/0112102.
  \bibitem{bijnens01} J. Bijnens, E. Pallante, and J. Prades,
    \Journal{\NPB}{626}{410}{2002}.
    \bibitem{musolf02} M. Ramsey-Musolf and Mark B. Wise,
 \Journal{\PRL}{89}{041601}{2002}.
\end{thebibliography}
\end{document}